\documentclass[nofootinbib,aps,amsmath,amssymb,a4paper,prl,twocolumn,preprintnumbers]{revtex4-1}
\usepackage[american]{babel}
\usepackage[utf8]{inputenc}
\usepackage[T1]{fontenc}
\usepackage{graphicx}
\usepackage{color}
\usepackage{units}
\usepackage{hyperref}
\hypersetup{
   pdfnewwindow=true,    
   colorlinks=true,      
   linkcolor=blue,       
   citecolor=blue,       
   filecolor=blue,       
   urlcolor=blue         
}

\begin{document}
\title{\Large {\bf{Gamma-Ray Excess and the Minimal Dark Matter Model}}}
\author{Michael Duerr,$^{1}$ Pavel Fileviez P\'erez,$^{2}$ Juri Smirnov$^{2}$}        
\affiliation{\vspace{3ex} $^{1}$DESY, Notkestra\ss{}e 85, D-22607 Hamburg, Germany \\
$^{2}$Particle and Astroparticle Physics Division, Max-Planck-Institut f\"ur Kernphysik, \\ Saupfercheckweg 1, 69117 Heidelberg, Germany } 
\preprint{DESY 15-191}
\begin{abstract}
We point out that the gamma-ray excesses in the galactic center and in the dwarf galaxy 
Reticulum II can both be well explained within the simplest dark matter model. 
We find that the corresponding regions of parameter space will be tested by direct and indirect dark matter searches in the near future. 
\end{abstract}
\maketitle
\section{Introduction}
The nature of the dark matter~(DM) in the Universe is one of the most fascinating mysteries of modern cosmology. Unfortunately, we know little about the properties 
of the dark matter apart from having a very good measurement of the relic density today, $\Omega_\text{DM} h^2 = 0.1199 \pm 0.0022$~\cite{Ade:2015xua}. Thanks to the efforts of the experimental community we also have several bounds from direct and indirect detection experiments, which are very important to constrain dark matter models.

The simplest dark matter model was proposed 30 years ago by Silveira and Zee~\cite{Silveira:1985rk}. In this context the dark matter candidate is a real scalar 
field which is stabilized by a discrete symmetry. This simple model can be considered as a toy model which allows to predict all the properties of the dark 
matter candidate once the relic density constraints are imposed. See Refs.~\cite{McDonald:1993ex,Burgess:2000yq,O'Connell:2006wi,Barger:2007im,Yaguna:2008hd,He:2008qm,Farina:2009ez,Guo:2010hq,Profumo:2010kp,Kadastik:2011aa,Djouadi:2012zc,Cheung:2012xb,Cline:2013gha,Khan:2014kba,Feng:2014vea,Kahlhoefer:2015jma,Duerr:2015mva,Han:2015hda,Duerr:2015aka} for previous studies of this model.

Recently, a possible gamma-ray excess coming from the center of our galaxy has been reported, and many studies by several groups lead to the general 
consensus that there is such an excess~\cite{Goodenough:2009gk,Hooper:2010mq,Boyarsky:2010dr,Hooper:2011ti,Linden:2012iv,Abazajian:2012pn,Hooper:2013rwa,Gordon:2013vta,Abazajian:2014fta,Daylan:2014rsa,Zhou:2014lva,Calore:2014xka,Calore:2014nla,Agrawal:2014oha,Porter:2015uaa}. A recent analysis by the Fermi-LAT collaboration~\cite{TheFermi-LAT:2015kwa} confirms a residual emission from the galactic center. One explanation for this excess could be dark matter annihilation. This possibility has been investigated by many groups, discussing the allowed region for the dark matter mass and the annihilation cross section needed to explain 
the excess. See, e.g., Refs.~\cite{Calore:2014xka,Calore:2014nla,Agrawal:2014oha} for recent analyses. If these results are verified by current or future experiments one could reveal some of the properties of the dark matter in our galaxy.

In this letter we investigate the possibility to account for the gamma-ray excess in the context of the simplest dark matter model. This possibility has been mentioned 
in previous studies where the authors concluded that it is not possible to explain the excess, see for example Ref.~\cite{Cline:2013gha}. The study in Ref.~\cite{Cline:2013gha} was done using only dark matter masses $\lesssim \unit[60]{GeV}$ and assuming annihilation into two bottom quarks, since at that time this was thought to be the only allowed mass region for the excess. 

Given the uncertainty in the background model~\cite{TheFermi-LAT:2015kwa}, of course no conclusive statement on the origin of the excess can be made. While the apparent excess might have some astrophysical origin not properly accounted for in the background model, the uncertainties also result in a wider range of DM masses and models being able to fit the excess. This has been explicitly discussed in, e.g., Refs.~\cite{Calore:2014nla,Agrawal:2014oha} and it is therefore worthwile to reconsider the scalar singlet DM model in some detail and check whether this simple model can provide an explanation for the excess.

In this letter we show that it is possible to account for the galactic center excess and the excess in Reticulum~II~\cite{Geringer-Sameth:2015lua} in two regions of parameter space in this minimal model: in the low mass region very close to the Standard Model~(SM) Higgs resonance where the dark matter annihilates mainly into bottom quarks and in the high mass region where the main annihilation channel is into two $W$ gauge bosons. 
We discuss the correlation between the gamma-ray excesses and the predictions for direct detection experiments and the gamma-ray lines in both mass regions.

\section{Minimal Dark Matter Model}
The minimal dark matter model~\cite{Silveira:1985rk} is the simplest extension of the Standard Model, since one adds 
only a real scalar field, $S \sim (1,1,0)$, to the Standard Model particle content. In this model the Higgs portal coupling defines both the relic density and the predictions for direct and indirect dark matter detection. 

The Lagrangian of this model reads as
 \begin{equation}
 \mathcal{L}=\mathcal{L}_\text{SM} + \frac{1}{2} \partial_\mu S \partial^\mu S -  \frac{1}{2} m_S^2 S^2 - \lambda_S S^4 - \lambda_p H^\dagger H S^2,
 \end{equation} 
where $H\sim (1,2,1/2)$ is the SM Higgs and $\mathcal{L}_\text{SM}$ is the usual SM Lagrangian. Notice that this model has only two relevant parameters for our study, the Higgs portal coupling $\lambda_p$ and the physical dark matter mass $M_S$. Here $M_S^2=m_S^2 + \lambda_p v_0^2$ where $v_0$ is the vacuum expectation value of the SM Higgs. A discrete 
$\mathcal{Z}_2$ symmetry is imposed to guarantee the stability of the dark matter candidate. Different aspects of this model have been investigated by many groups, see, e.g., Refs.~\cite{McDonald:1993ex,Burgess:2000yq,O'Connell:2006wi,Barger:2007im,Yaguna:2008hd,He:2008qm,Farina:2009ez,Guo:2010hq,Profumo:2010kp,Kadastik:2011aa,Djouadi:2012zc,Cheung:2012xb,Cline:2013gha,Khan:2014kba,Feng:2014vea,Kahlhoefer:2015jma,Duerr:2015mva,Han:2015hda,Duerr:2015aka}.

\begin{figure}[t]
 \includegraphics[width=0.75\linewidth]{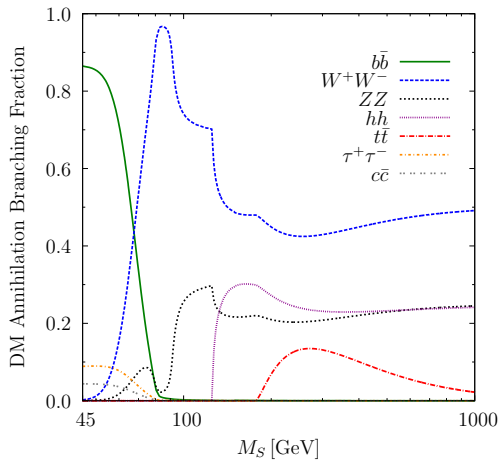}
 \caption{Branching fractions for the various DM annihilation channels in agreement with the relic density.}
 \label{fig:BR}
\end{figure}

In order to facilitate the discussion in the next sections we show in Fig.~\ref{fig:BR} the branching ratios for the different annihilation channels in agreement with the full relic density. In the low mass region the main annihilation channel is into two bottom quarks, while in the heavy region the annihilation into two $W$ gauge bosons dominates.
This is crucial to understand the possibility to explain the gamma-ray excess from the center of the galaxy, which we discuss in the next section.

\section{Galactic Center Excess}

\begin{figure}[t]
 \includegraphics[width=0.8\linewidth]{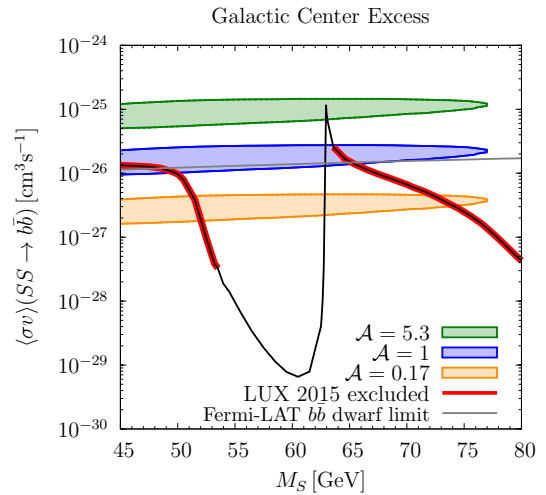}
 \caption{Regions in agreement with the excess in the galactic center (colored) and the DM annihilation cross section (black solid) in the minimal model when the dark matter mass is below the gauge boson masses. The dominant channel then is the annihilation into $b$ quarks, $SS \to b \bar{b}$. 
 The colored regions are at the 3$\sigma$ level, and we use three values for $\mathcal{A}$, which parametrizes the variation in the dark matter profile~\cite{Calore:2014nla}.
 The red part of the curve is excluded by LUX~\cite{Akerib:2015rjg}, and the gray line corresponds to the limit from the Fermi-LAT collaboration~\cite{Ackermann:2015zua}.}
 \label{fig:centerbbbar}
\end{figure}

\begin{figure}[t]
 \includegraphics[width=0.8\linewidth]{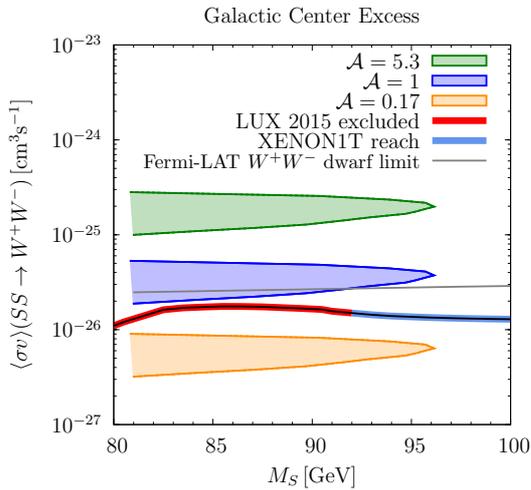}
 \caption{Regions in agreement with the excess in the galactic center (colored) and the annihilation cross section (black solid) in the minimal model when the dark matter mass is larger than the $W$ gauge boson mass. The dominant channel then is the annihilation $SS \to WW$. 
 The colored regions are at the 3$\sigma$ level, and we use three values for $\mathcal{A}$, which parametrizes the variation in the dark matter profile~\cite{Calore:2014nla}.
 The red part of the curve is excluded by LUX~\cite{Akerib:2015rjg}, and the blue part of the curve shows the future reach of XENON1T~\cite{Aprile:2015uzo}. The gray line corresponds to the limit from the Fermi-LAT collaboration~\cite{Ackermann:2015zua}.}
 \label{fig:centerWW}
\end{figure}

The existence of a possible gamma-ray excess in the center of the galaxy has been investigated by many groups~\cite{Goodenough:2009gk,Hooper:2010mq,Boyarsky:2010dr,Hooper:2011ti,Linden:2012iv,Abazajian:2012pn,Hooper:2013rwa,Gordon:2013vta,Abazajian:2014fta,Daylan:2014rsa,Zhou:2014lva,Calore:2014xka,Calore:2014nla,Agrawal:2014oha,Porter:2015uaa,TheFermi-LAT:2015kwa}, and the general consensus 
is that one could explain the excess via the annihilation of dark matter. Here we use the numerical results presented in Ref.~\cite{Calore:2014nla} 
to constrain the simplest dark matter model.

In Fig.~\ref{fig:centerbbbar} we show the allowed parameter space in the low mass region where the main annihilation channel is $SS \to b \bar{b}$.
We use the range $\mathcal{A}=0.17-5.3$ to parametrize the uncertainties in the dark matter halo of the Milky Way~\cite{Calore:2014nla}. In red we show 
the region excluded by the LUX experiment~\cite{Akerib:2015rjg} and the gray line shows the limit for the annihilation into $b\bar{b}$ from Fermi-LAT~\cite{Ackermann:2015zua}.
Notice that there is a small region with the dark matter mass in the range $M_S=\unit[(62-63)]{GeV}$ which is allowed by the experiments and one can explain the excess in the galactic center. 

The main features of this region with the dark matter in the range $M_S=\unit[(62-63)]{GeV}$ are:
\begin{itemize}
\item The cross section $SS \to \gamma \gamma$ is very close to the experimental limit set by Fermi-LAT~\cite{Ackermann:2015lka}; see Fig.~\ref{fig:PredictionsGammaLines} for more details. Therefore, this region will be tested soon.
\item The relic density is set using the Higgs resonance.
\end{itemize}
Unfortunately, in this region the spin-independent DM--nucleon cross section is very small and one cannot test this part of the parameter space at future direct detection experiments such as XENON1T~\cite{Aprile:2015uzo}. 

In the heavy dark matter region the main annihilation channel is $SS \to WW$, and one can explain the excess in the galactic center in agreement with LUX~\cite{Akerib:2015rjg} if the dark matter mass is in the range $M_S=\unit[(92-96)]{GeV}$, as we show in Fig.~\ref{fig:centerWW}. Notice that the allowed region is larger than in the low mass scenario.

This region has two main features:
\begin{itemize}
\item The spin-independent DM--nucleon cross section is large and this region can be tested or excluded by XENON1T~\cite{Aprile:2015uzo}; see the blue line in Fig.~\ref{fig:centerWW}.
\item The main annihilation channel is $WW$ where the branching ratio changes between $90\%-95\%$. This scenario is ideal to use the results from Ref.~\cite{Calore:2014nla}.
\end{itemize}
Unfortunately, in the heavy mass region it is challenging to test this model at gamma-ray telescopes soon because the annihilation cross section into photons is small; see Ref.~\cite{Duerr:2015aka} and Fig.~\ref{fig:PredictionsGammaLines} for details.

In summary, in the minimal dark matter model one can have two consistent scenarios to explain the galactic center gamma-ray excess in agreement with all experimental constraints.

\section{Predictions for the Gamma-Ray Lines}

\begin{figure}[t]
 \includegraphics[width=0.8\linewidth]{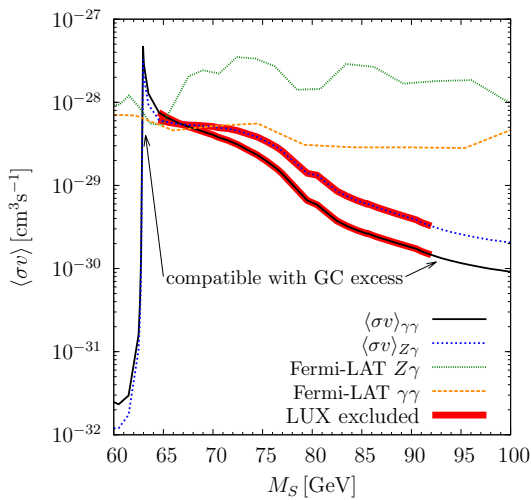}
 \caption{Annihilation into gamma-ray lines in agreement with the relic density. Parts of the curve in the low mass and the high mass region are in agreement with the gamma-ray excess in the center of the galaxy.}
 \label{fig:PredictionsGammaLines}
\end{figure}

Using the two regions compatible with the gamma-ray excess from the galactic center, which we discussed above, we show the predictions for the gamma-ray lines in this model in Fig.~\ref{fig:PredictionsGammaLines}. In the low mass scenario the cross section is very close to the Fermi-LAT limit~\cite{Ackermann:2015lka}. In the heavy mass region the annihilation cross section into two photons is almost two orders of magnitude smaller. 

Now, let us discuss the visibility of these gamma-ray lines. Recently, we have investigated this issue in detail~\cite{Duerr:2015mva,Duerr:2015aka}, pointing out the role of the final state radiation processes. 

\begin{figure}[t]
 \includegraphics[width=0.8\linewidth]{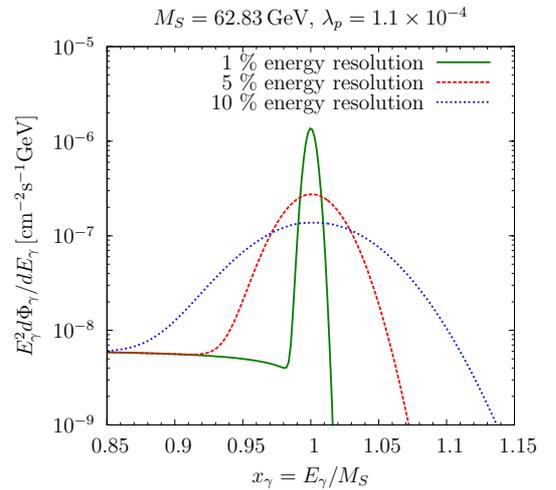}
 \caption{Gamma-ray spectrum for a dark matter mass of $M_S=\unit[62.83]{GeV}$. The portal coupling $\lambda_p$ is chosen to reproduce the correct relic density. 
 For this value of the DM mass, the position of the $Z\gamma$ line is at $\unit[30]{GeV}$. The main contribution to the continuum close to the line is coming from the final state radiation process $SS \to b\bar{b} \gamma$.}
 \label{fig:visibilityLow}
\end{figure}

In Fig.~\ref{fig:visibilityLow} we show the gamma-ray spectrum for a dark matter mass of $M_S=\unit[62.83]{GeV}$, being in the low mass region. 
Then, the main contribution to the continuum close to the gamma-ray line comes from the final state radiation process $SS \to b\bar{b}\gamma$. See Ref.~\cite{Duerr:2015aka} for the full analytical results. In Fig.~\ref{fig:visibilityLow} one can appreciate that this gamma-ray line could be easily seen because the final state 
radiation is suppressed by the small bottom Yukawa coupling. Since the prediction for the cross section for this gamma-ray line is close to Fermi-LAT limit one can test this scenario in the near future.

\begin{figure}[t]
 \includegraphics[width=0.8\linewidth]{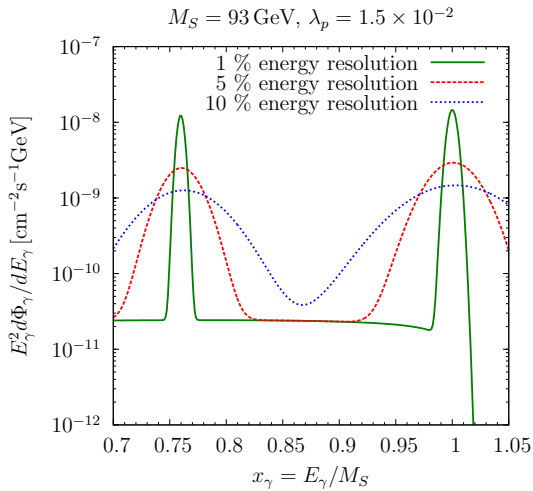}
 \caption{Gamma-ray spectrum for a dark matter mass of $M_S=\unit[93]{GeV}$. The portal coupling $\lambda_p$ is chosen to reproduce the correct relic density.  
The main contribution to the continuum close to the line is coming from the final state radiation process $SS \to WW \gamma$. }
 \label{fig:visibilityHigh}
\end{figure}

In Fig.~\ref{fig:visibilityHigh} we show the gamma-ray spectrum for the heavy mass scenario, for $M_S=\unit[93]{GeV}$.
The main contribution to final state radiation in this case is due to the dark matter annihilation into $WW\gamma$. As one can see, 
in this scenario one could see both lines, $\gamma \gamma$ and $Z \gamma$, due to the fact that there is a large difference 
between the continuum and the lines. Unfortunately, in this case the annihilation cross sections to gamma-ray lines are quite below the current Fermi-LAT limit~\cite{Ackermann:2015lka}.

\section{Gamma-Ray Excess and Reticulum II}
Recently, a possible gamma-ray excess has been reported in the dwarf galaxy Reticulum II~\cite{Geringer-Sameth:2015lua}; see also Refs.~\cite{Drlica-Wagner:2015xua,Hooper:2015ula}. 
The energy and intensity of the excess is such that it can be possibly related to the excess from the galactic center. Here we show that it is possible to have a consistent 
scenario in the minimal dark matter model in both mass ranges discussed above. 

\begin{figure}[t]
 \includegraphics[width=0.8\linewidth]{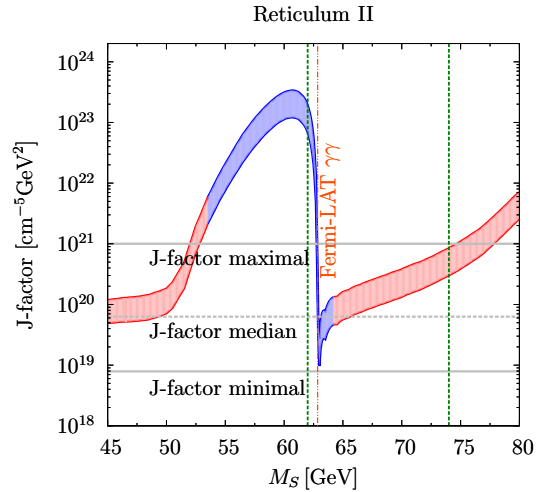}
 \caption{Predictions for the $J$-factor of Reticulum II in the low mass region, assuming a dark matter origin of the excess. The maximal, minimal, and median value for the $J$-factor are taken from Ref.~\cite{Bonnivard:2015tta}. The region between the two green dashed lines is in agreement with the galactic center excess in this model. The red dash-dotted line corresponds to the $\gamma \gamma$ limit from Fermi-LAT~\cite{Ackermann:2015lka} which excludes the region to the right. The regions in red are excluded and the regions in blue are allowed by the LUX experiment.}
 \label{fig:ReticulumLow}
\end{figure}

\begin{figure}[t]
 \includegraphics[width=0.8\linewidth]{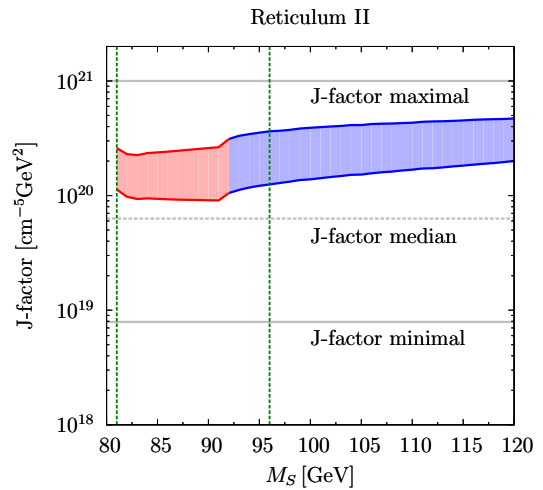}
 \caption{Same as Fig.~\ref{fig:ReticulumLow} for the high mass region.}
 \label{fig:ReticulumHigh}
\end{figure}

In Figs.~\ref{fig:ReticulumLow} and \ref{fig:ReticulumHigh} we show the allowed values for the $J$-factor if this gamma-ray excess is explained by dark matter annihilation. In the low mass region the annihilation is predominantly to $b\bar{b}$ and in the high mass region to $WW$. The values for the $J$-factor we obtain 
are around $\unit[10^{20}]{GeV^2 cm^{-5}}$ in both cases. The horizontal lines show the minimal, maximal and median values for the $J$-factor for Reticulum II~\cite{Bonnivard:2015tta}. In Fig.~\ref{fig:ReticulumHigh} we find excellent agreement of the predicted $J$-factor with the estimates in Ref.~\cite{Bonnivard:2015tta} in the full mass range. 

In Fig.~\ref{fig:ReticulumLow} we show the $J$-factor for the low mass regime, and it turns out that it is compatible with Ref.~\cite{Bonnivard:2015tta} in the case the DM mass is above $\unit[62]{GeV}$. Thus, there is a strong pull by the data from Reticulum II towards mass values in which the annihilation signal to gamma gamma is strong, see Fig.~\ref{fig:PredictionsGammaLines}. 

Therefore, with this example we illustrate the possibility to explain both the gamma-ray excess in Reticulum~II and the excess in the galactic center in the context of the minimal dark matter model.

\section{Summary}
In this letter we have shown that in the context of the minimal dark matter model one can have two consistent scenarios where one can explain the gamma-ray excess in the galactic center, in agreement with all other experimental constraints. 

In the low mass region the dark matter mass is very close to half the Standard Model Higgs mass and through the resonance one can explain the relic density. In this case the main annihilation is $SS \to b \bar{b}$ and the predictions for the cross section $SS \to \gamma \gamma$ are very close to the experimental limits. In the heavy mass region, for $M_S=\unit[(92-96)]{GeV}$, the dark matter candidate annihilates mainly into two $W$ gauge bosons. The interesting feature of this region is that the predictions for the spin-independent nucleon--DM cross section are in the region that can be tested by the XENON1T experiment.

We have also shown the allowed values for the $J$-factor for Reticulum II in order to understand the correlation between the gamma-ray excess in the center of the galaxy and the excess in Reticulum II. 

Our main conclusion is that both gamma-ray excesses can be explained within the minimal model for dark matter and there is no need to do model building to accommodate these signals.

\section{Acknowledgments} 
We would like to thank A.\ Geringer-Sameth for helpful communication and for providing us with the fit data for the $WW$ channel and F.\ S.\ Queiroz for discussions. 
M.\ D.\ is supported by the German Science Foundation (DFG) under the Collaborative Research Center (SFB) 676 Particles, Strings and the Early Universe as well as the ERC Starting Grant `NewAve' (638528).


\end{document}